# Status of the Bio-Nano electron cyclotron resonance ion source at Toyo University[a]

T. Uchida,[1,b] H. Minezaki,[2] S. Ishihara,[2] M. Muramatsu,[3] R. Rácz,[4] T. Asaji,[5] A. Kitagawa,[3] Y. Kato,[6] S. Biri,[4] A. G. Drentje,[3] and Y. Yoshida[1,2]

[1]*Bio-Nano Electronics Research Centre, Toyo University, Kawagoe 350-8585, Japan*

[2]*Graduate School of Engineering, Toyo University, Kawagoe 350-8585, Japan*

[3]*National Institute of Radiological Sciences (NIRS), Chiba 263-8555, Japan*

[4]*Institute for Nuclear Research (ATOMKI), H-4026 Debrecen, Hungary*

[5]*Oshima National College of Maritime Technology, Yamaguchi 742-2193, Japan*

[6]*Graduate School of Engineering, Osaka University, Suita 565-0871, Japan*

In the paper, the material science experiments, carried out recently using the Bio-Nano electron cyclotron resonance ion source (ECRIS) at Toyo University, are reported. We have investigated several methods to synthesize endohedral $C_{60}$ using ion-ion and ion-molecule collision reaction in the ECRIS.
Because of the simplicity of the configuration, we can install a large choice of additional equipment in the ECRIS. The Bio-Nano ECRIS is suitable not only to test the materials production but also to test technical developments to improve or understand the performance of an ECRIS.

## I. INTRODUCTION

Endohedral fullerene, one of the carbon nanostructures, is a fullerene derivative where an alien atom(s) resides in the fullerene cage. The alien atom can affect the material character. Therefore, for example, endohedral nitrogen-$C_{60}$ (N@$C_{60}$), endohedral argon-$C_{60}$, endohedral gadolinium-$C_{60}$, and endohedral lithium-$C_{60}$ (Li@$C_{60}$) are thought to be promising materials in applications of quantum computing, superconductor, magnetic resonance imaging contrast agent, and molecular switch, respectively.[1–4] However, these materials are seldom used in industrial applications. One of the main reasons for that is the production of these materials. The endohedral fullerenes can be produced by collision reaction of the fullerene molecules with the encapsulating atoms (collision processing). The collision processing can be taken place in the high-pressure/high-temperature vapor phase,[5] in the plasma,[6] and by particle-surface interaction using an ion beam (ion irradiation technique).[7] Among the production methods of endohedral fullerenes, the ion irradiation technique is the most effective method in terms of the yield so far. Furthermore, the production processes, including the synthesis and purification, can be simplified.

Two of the authors in this paper have synthesized N@$C_{60}$ using nitrogen-$C_{60}$ mixture plasma in an ECRIS.[6] The N@$C_{60}$ could be observed in the extracted ion beam with the ion current of less than 1 pnA. The advantage of this method is that the synthesized endohedral fullerenes can be separated from the parent fullerene molecules and other byproducts depending on the m/q ratio. Though the separation of the endohedral ones from the exohedral ones is not possible. Based on the previous report, we have developed an ECRIS at Toyo University (the Bio-Nano ECRIS), which is designed for the production of endohedral fullerenes, in particular, endohedral iron-$C_{60}$ (Fe@$C_{60}$), by the collision processing, since 2006.[8,9] With our special ion source, one can carry out investigations in different ways. They are: (1) Synthesis of $C_{60}$ and X in the plasma by ion-ion, ion-neutral, and neutral-neutral collisions. The charge of the ions can be positive or negative. (2) Synthesis at the border of the plasma, on a biased, water-cooled electrode. The two components (ions or atoms) arrive from the plasma. (3) Synthesis on a beamline target. Pre-prepared $C_{60}$ layers are irradiated with slow X ions. (4) Opposite than 4. Pre-prepared X-layer is irradiated with slow $C_{60}$ ions. (5) A continuously depositing $C_{60}$-layer is irradiated with X ions. We have tested several methods to synthesize endohedral fullerenes in the ECRIS; one-chamber and two-chambers configurations. In the one-chamber configuration, we have investigated the generation of a pure $C_{60}$ plasma using gas mixing and RF pulse modulation techniques,[10] and also the generation of a pure iron plasma using an induction-heating oven.[9] In the two-chambers configuration, we have investigated the fundamental characteristics, such as plasma generation in the gas-injection side and/or the extraction side chambers.[11] Also, in order to study the feasibility of the Fe@$C_{60}$ synthesis, we have performed the $Fe^+$ irradiation to $C_{60}$ layers and could synthesize an Fe-$C_{60}$ complex, which might be the Fe@$C_{60}$.[12,13]

In this paper, we report our recent activities on the synthesis of endohedral fullerenes using the one-chamber and two-chambers configurations of the Bio-Nano ECRIS at Toyo University.

## II. EXPERIMENTAL SETUP

There are several device configurations in the Bio-Nano ECRIS apparatus: one-chamber, and two-chambers

---

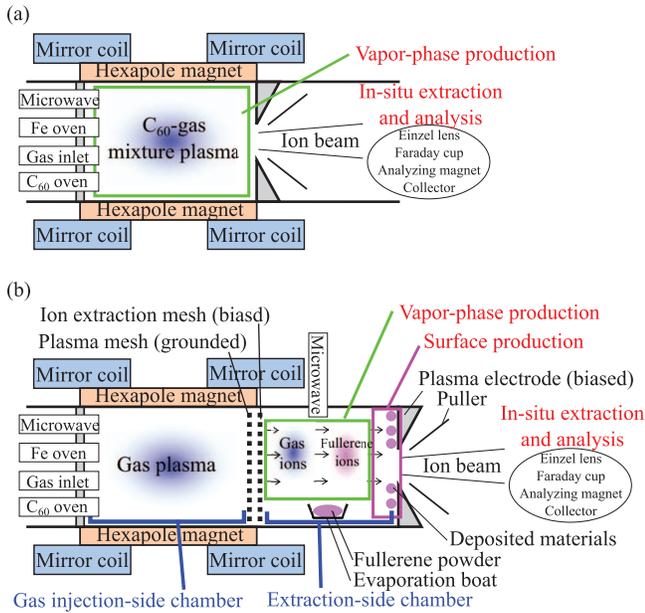

FIG. 1. Schematic drawings of (a) one-chamber and (b) two-chambers configurations of the Bio-Nano ECRIS.

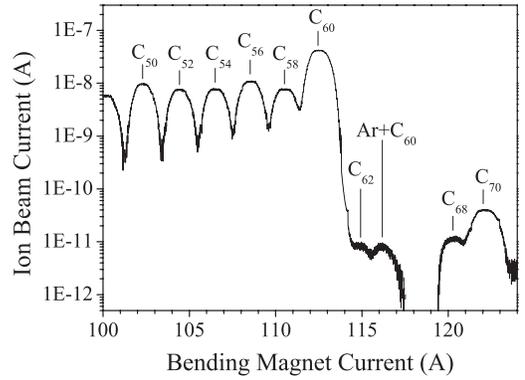

FIG. 2. Beam spectrum of extracted ions from the mixture plasma of $C_{60}$ and Ar in the ECRIS.

configurations (Fig. 1). The one-chamber configuration is a normal ECRIS with min-B configuration. This is mainly used to perform the synthesis of endohedral fullerene by the ion-ion or ion-neutral collision reaction in the mixture plasma (vapor-phase production) of the ECRIS or by the ion irradiation in the beamline of the ECRIS. Using the beamline, the ions with m/q = 2000 at an extraction voltage of 5 kV can be transported. In the two-chambers configuration, in contrast to the one-chamber configuration, an additional plasma chamber is added to the installed pair of metal mesh electrodes. Both in the injection-side and extraction-side chambers, there are individual 10 GHz microwave supplies. By applying negative bias voltage to the mesh electrode, we can extract ions from the injection-side chamber to the extraction-side one. Also, the plasma electrode can be biased negatively in order to irradiate ions to the substrate at a specific energy. Therefore, in the two-chambers configuration, we can perform the two synthesis methods simultaneously: a vapor-phase production by extracted ions from the injection-side chamber and fullerene ions or neutrals made in the extraction-side chamber, and a surface collision production at the plasma electrode. The details of the device configurations are described in Refs. 8 and 11.

## III. RESULTS AND DISCUSSION

### A. One-chamber configuration

In order to produce, separate and collect the endohedral fullerenes using an ECRIS and the following beamline, there has been several challenges: (1) to make the endohedral fullerene ions with high ion beam intensity, (2) to separate the endohedral fullerene ions with the other ones, and (3) to collect the beam of the endohedral fullerene ions with high kinetic energy (in our case, the energy is 5 keV) without destructing the structure. Regarding the challenges (1) and (2), we have investigated the synthesis using several mixture plasmas: N-$C_{60}$, Fe-$C_{60}$, and Ar-$C_{60}$. First, we used N-$C_{60}$ mixture plasma to reproduce the previous result. But it was difficult to detect the peak of the N@$C_{60}$ in the beam spectrum, which may be caused by the insufficiency of the ion beam separation resolution. As we can see in Fig. 2, the foot of the $C_{60}$ peak reaches the mass of N + $C_{60}$. Then, we shifted to use Ar-$C_{60}$ mixture plasma, in which Ar + $C_{60}$ has larger mass difference from $C_{60}$ than N + $C_{60}$. Fig. 2 shows the beam spectrum of extracted ions from the mixture plasma of $C_{60}$ and Ar in the ECRIS, which is optimized to get the highest beam intensity of Ar + $C_{60}$. The Ar flow rate, chamber pressure, $C_{60}$ oven temperature, microwave frequency, and microwave power were, respectively, 0.19 sccm, $1 \times 10^{-4}$ Pa, 450 °C, 9.75 GHz, and 1.8 W. The peak with a mass corresponds to Ar + $C_{60}$ was observed together with the fullerenes such as $C_{70}$, $C_{68}$, $C_{62}$, $C_{60}$, $C_{58}$, …. The intensity of Ar + $C_{60}$ ions was only 10 pA and much lower than that of $C_{60}$. Regarding the challenge (3), we have designed and tested the beam deceleration system at the end of the beamline. Using the beam deceleration system, we can collect $C_{60}^{+}$ ions as $C_{60}$ layers on a substrate without destructing the $C_{60}$ cage structure.[6] There still remains the challenges (1) and (2). In order to increase the beam intensity of the ions with a mass corresponds to en-dohedral fullerenes, we will further test experimental parameters, in particular, such as chamber pressure. Also, we must consider how to separate endohedral fullerenes and exohedral ones. However, the separation of them using an off-line separation method like HPLC is not so difficult. We have developed an induction-heating oven for iron vapor supply in order to perform the synthesis of the endohedral Fe-$C_{60}$ (Fe@$C_{60}$) in the mixture plasma of $C_{60}$ and Fe. However, we did not succeed in the production of the pure Fe-$C_{60}$ mixture plasma and the synthesis of the Fe@$C_{60}$.[9]

### B. Two-chambers configuration

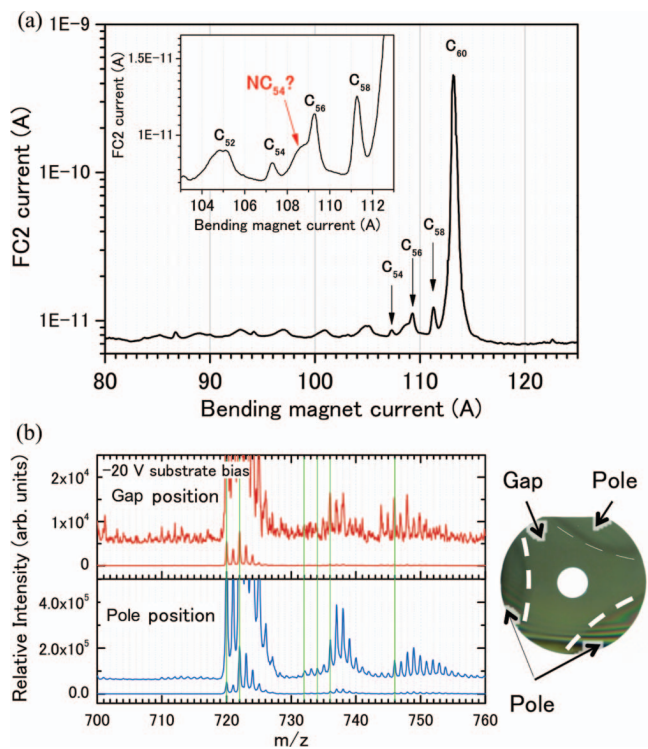

FIG. 3. (a) Beam spectrum of extracted ions and (b) LDI-TOF-MS spectra of the deposited material on the plasma electrode (substrate) of the gap and pole positions, obtained with $-20$ V bias voltage. The right-side photo shows the irradiated substrate on which materials were deposited. The gap and pole positions correspond to the region, respectively, inside or outside the triangular mark.

and extraction-side chambers. The $N_2$ flow rate, chamber pressure, evaporation boat temperature, microwave frequencies, and microwave powers were, respectively, 0.18 sccm, $5 \times 10^{-5}$ Pa, 400 °C, 9.75 GHz (gas-injection side), 9.9 GHz (extraction-side), 40 W (gas-injection side), and 8 W (extraction-side). In the extracted ion beam spectrum, we could observe a new peak between $C_{56}^+$ and $C_{54}^+$, which can correspond to $N + C_{54}^+$ [Fig. 3(a)]. From the plasma electrode, we could obtain $C_{60}$ molecules incorporating nitrogen, carbon, and oxygen, such as $NC_{59}$ (m/q = 722), $C_{61}$ (m/q = 732), $N + C_{60}$ (m/q = 734), $O + C_{60}$ (m/q = 736), $N + C_{61}$ (m/q = 746), $O + C_{61}$, (m/q = 748), etc. [Fig. 3(b)]. In this experiment, gas sources were only nitrogen and $C_{60}$, but there were impurities of carbon and oxygen. Beam current ratio is as follows; $C^+ : N^+ : O^+ = 1 : 3 : 1$. That is because we observed carbon or oxygen-incorporating $C_{60}$ materials. The intensities of the above peaks depend on the positions in the substrate and the bias voltage. From the pole position, where the electrons and ions were hard to reach because of the magnetic field of the hexapole, we could always get higher intensity of the above peaks than those from the gap position. This indicates that the charged particles, mainly electrons, damage the deposited materials. Also the intensities of the above peaks decrease with an increase of the substrate bias. The $-10$ V bias voltage, which was the lowest bias voltage in the experiments, showed the highest intensities.

## IV. SUMMARY

We reported the investigation of the synthesis of endohedral $C_{60}$ using the Bio-Nano ECRIS. We could observe the ions of $Ar + C_{60}$ in the beam spectrum using the one-chamber configuration, and could observe the $C_{60}$ molecules incorporating nitrogen, carbon, and oxygen using the two-chambers configuration. Also, we would emphasize that we can install a large choice of additional equipment in the ECRIS, because of the simplicity of the configuration of the Bio-Nano ECRIS (e.g., see Ref. 14). The Bio-Nano ECRIS is suitable not only to test materials production but also to test technical developments to improve or understand the performance of an ECRIS.

## ACKNOWLEDGMENTS

Part of this study has been supported by JSPS KAKENHI Grant Nos. 24810029, 24710095, and a Grant for the Strategic Development of Advanced Science and Technology S1101017 organized by the Ministry of Education, Culture, Sports, Science, and Technology, Japan. The participation of two of the authors (S.B. and R.R.) in this work was partly supported by the TAMOP 4.2.2.A-11/1/KONV-2012-0036 project, which is co-financed by the European Union and European Social Fund.